\documentclass[final,3p,times,twocolumn]{elsarticle}

\usepackage{textcomp}

 \usepackage{graphics}
 \usepackage{graphicx}
\usepackage{amssymb}
\usepackage{lineno}
\journal{Nuclear Instruments and Methods  A}

\begin{document}

\begin{frontmatter}



\title{A Multi-Channel THz and Infrared Spectrometer for Femtosecond Electron Bunch Diagnostics by Single-Shot Spectroscopy of Coherent Radiation}


\author[desy]{Stephan~Wesch}
\author[desy]{Bernhard~Schmidt}
\author[desy]{Christopher~Behrens}
\author[desy]{Hossein~Delsim-Hashemi}
\author[desy]{Peter~Schm\"user}

\address[desy]{Deutsches Elektronen-Synchrotron DESY, Notkestra\ss{}e 85, 22607 Hamburg, Germany}

\begin{abstract}
The required high peak current in free-electron lasers (FELs) is realized by longitudinal compression of the electron bunches to sub-picosecond length. In this paper, a frequency-domain  diagnostic method is described that is capable of resolving structures in the femtosecond regime. A novel in-vacuum  spectrometer has been developed for spectroscopy of coherent radiation in the THz and infrared range. The spectrometer is equipped with five consecutive  dispersion gratings and 120 parallel readout channels; it can be operated either in short (5~-~44\,\textmu m) or in long wavelength mode (45~-~430\,\textmu m). Fast parallel readout  permits the spectroscopy of coherent radiation from single electron bunches. Test measurements at the soft X-ray free-electron laser FLASH, using  coherent  transition radiation, demonstrate excellent performance of the spectrometer. The device is planned for use as an online bunch profile monitor during regular FEL operation.
\end{abstract}


\end{frontmatter}

% \linenumbers

\section{Introduction}
\label{intro}

The electron bunches in high-gain free-electron lasers are longitudinally compressed to achieve peak currents in the kA range which are necessary to drive the FEL gain process in the undulator magnets. Bunch compression is accomplished by a two-stage process: first an {\it energy chirp} (energy-position relationship) is imprinted onto the typically 10\,ps long bunches emerging from the electron gun, and then the chirped bunches are passed through magnetic chicanes.

Magnetic compression of intense electron bunches is strongly affected by collective effects in the chicanes and cannot be adequately described by linear beam transfer theory. Space charge forces and coherent synchrotron radiation have a profound influence on the time profile and internal energy distribution of the compressed bunches. The collective effects have been studied by various numerical simulations (see~\cite{Dohlus} and the references quoted therein) but the parameter uncertainties are considerable and experimental data are thus indispensable for determining the longitudinal bunch structure. Using a transverse deflecting microwave structure (TDS)~\cite{TDS1,TDS2} our group has carried out a time-resolved phase space tomography~\cite{Roehrs} of the compressed bunches in the free-electron laser FLASH. The TDS converts the temporal profile of the electron bunch charge density into a transverse streak on a view screen by a rapidly varying electromagnetic field, analogous to the sawtooth voltage in a conventional oscilloscope tube. The time resolution depends on the microwave voltage and the beam optics in the diagnostic section. In the present beam optics configuration at FLASH, a resolution of down to 25\,fs rms has been achieved. 

The electro-optical (EO) technique is an alternative method of measuring the longitudial bunch charge distribution. Several variants of single-shot EO bunch diagnostics have been applied~\cite{wilke_2002,berden_2004,cavalieri_2005}, all sharing the underlying principle of utilizing the field-induced birefringence in an electro-optic crystal to convert the time profile of a bunch into a temporal, spectral, or spatial intensity modulation of a probe laser pulse. The EO techniques have the advantage of being non-destructive, thereby permitting correlation studies of EO measurements on selected bunches with the FEL radiation pulses produced by the same bunches. In Ref.~\cite{berden_2007} an absolute calibration of the EO technique was performed utilizing simultaneous TDS measurements. A record resolution of 50\,fs (rms) in the detection of single electron bunches was achieved.
 
Frequency-domain techniques provide a complementary access to the femtosecond time regime. The spectral intensity of the coherent radiation emitted by a bunch with $N$ electrons is
	\begin{equation}
	 \frac{dU}{d\lambda}=\frac{dU_1}{d\lambda} \, N^2 \, \left| F_{\ell}(\lambda) \right|^2
	 \label{eq1}
	\end{equation}
where $ dU_1/d\lambda$ is the intensity per unit wavelength emitted by a single electron,  and $F_{\ell}(\lambda)$ is the longitudinal form factor of the bunch, the Fourier transform  of the normalized line charge density $S(z)$:  
	\begin{equation}
	 F_{\ell}(\lambda)=\int S(z) \, \exp(-2\pi i \, z/\lambda) \, dz
	\label{eq2}
	\end{equation}
In Eq.~(\ref{eq1}) we have made use of the fact that radiation from relativistic particles is confined to small angles whereby the influence of the transverse form factor is strongly suppressed. For an rms beam radius of 100\,\textmu m to 200\,\textmu m, as in our case, the determination of the longitudinal charge profile from the measured coherent radiation spectrum depends only weakly on the precise knowledge of the transverse charge distribution.

A measurement of the coherent radiation spectrum yields the absolute magnitude of the form factor as a function of wavelength but the phase remains unknown. Hence the determination of the longitudinal charge distribution by inverse Fourier transformation is not directly possible. Phase information can be obtained with the help of the Kramers-Kronig relation~\cite{Sievers} if a sufficient wavelength range is covered.

The typical lengths of compressed electron bunches of a few hundred micrometer leads to coherent emission in the far infrared (FIR) and millimeter wavelength range. The commonly used technique of spectroscopy for this regime is to measure the autocorrelation function with a Michelson type interferometer and to deduce the power spectrum by Fourier transformation (Fourier-spectroscopy). This technique has been applied for bunch length measurements at various electron linacs~\cite{Lai,Murokh,Geitz,Froehlich,Evtushenko,Thurman}. Since the interferometer is a step-scan device, it cannot be used for online monitoring. For this, polychromators measuring different wavelengths simultaneously have been proposed~\cite{Watanabe} but were of limited use due to their small number of read-out channels and thus restricted wavelength range.

We have developed a novel broadband spectroscopic instrument with single-shot capability~\cite{Hossein}. With two sets of five consecutive gratings, which can be interchanged by remote control, the most recent version of the spectrometer covers the far-infrared wavelength range from 45 to 435\,\textmu m or the mid-infrared range from 5 to 44\,\textmu m. The spectral intensity is recorded simultaneously in 120 wavelength bins. 

In this paper, we describe the design of the spectrometer, the detectors, amplifiers and readout electronics. Test measurements with coherent transition radiation (CTR) are presented which were carried out on bunches whose time profile was determined simultaneously using the transverse deflecting microwave structure TDS.

\section{Design and realization of the  spectrometer}
\label{spectrometer}

\subsection{Reflection gratings}
Coherent radiation from short electron bunches extends over a wide range in wavelength, from a few micrometers up to about 1\,mm.  Gratings are useful to disperse the polychromatic radiation into its spectral components. For a monochromatic plane wave that is  incident normally on a transmission grating with slit spacing $d$, the transmitted wave has interference maxima at angles $\beta_m$ given by the equation $d\sin{\beta_m}=m\,\lambda$. Here $m$ is the order of diffraction. The free spectral range of a grating is defined by the requirement that different diffraction orders do not overlap. Since light of wavelength $\lambda$, diffracted in the order $m=1$, will coincide with light of wavelength $\lambda/2$, diffracted in the order $m=2$, the ratio of the longest and the shortest wavelength in the free spectral range is close to two. Hence a single grating can only cover a small part of the entire coherent radiation spectrum, and many gratings of different spectral ranges have to be used simultaneously. Ambiguities due to overlap of different orders are avoided if the radiation impinging on a grating is filtered to limit the bandwidth appropriately. It will be shown below that this filtering is accomplished by a preceding grating. 

	\begin{figure}[htbp]
	 \centering
	 \includegraphics[width=0.7\columnwidth]{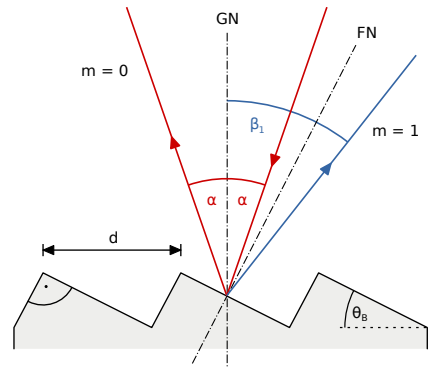}
	 \caption{Principle of a blazed reflection grating. For optimum efficiency of first-order diffraction, the incoming ray and the first-order diffracted ray have to enclose equal angles with respect to the facet normal $FN$. The blaze angle chosen is $\theta_B=27^\circ$, the incidence angle is $\alpha=19^\circ$ as measured with respect to the grating normal GN.}
	 \label{blazed-grating}
	\end{figure}

A transmission grating with a large number of slits, whose width is small compared to the slit separation, distributes the diffracted radiation power almost equally among  many orders. A great improvement in efficiency for a specific order is obtained using a blazed reflection grating~\cite{grating-handbook} with triangular grooves as shown in Fig.~\ref{blazed-grating}. For an incoming wave that is incident at an angle $\alpha$ with respect to the normal of the grating, the grating equation becomes 
	\begin{equation}\label{grating-equation}
	  d \, (\sin{\alpha} + \sin{\beta_m}) = m \,\lambda
	\end{equation}
To enhance the intensity in a given order $m$, the blaze angle $\theta_B$ and the incidence angle $\alpha$ are chosen such that the direction of the diffracted light coincides with the direction of specular reflection at the inclined surfaces~\cite{grating-handbook}. For the first order $m=1$ this implies $\theta_B-\alpha=\beta_1-\theta_B$, hence $\theta_B=(\alpha+\beta_1)/2$. In this case most of the diffracted power goes into the order $m=1$. 

It is a general feature of gratings that the diffraction effects vanish if the wavelength becomes too large. The incidence angle is $\alpha=19^\circ$ in our spectrometer setup, hence the largest possible value of $\sin{\alpha} + \sin{\beta_m} $ is 1.33. This implies that for wavelengths $\lambda \ge 1.33 \,d$, the grating equation~(\ref{grating-equation}) can only be satisfied with $m=0$ which means that no diffracted wave exists and the reflection grating behaves as a plane mirror. This specular reflection of long wavelengths is utilized in the multistage spectrometer described below.

	\begin{figure}[hb]
	 \centering
	 \includegraphics[width=0.85\columnwidth]{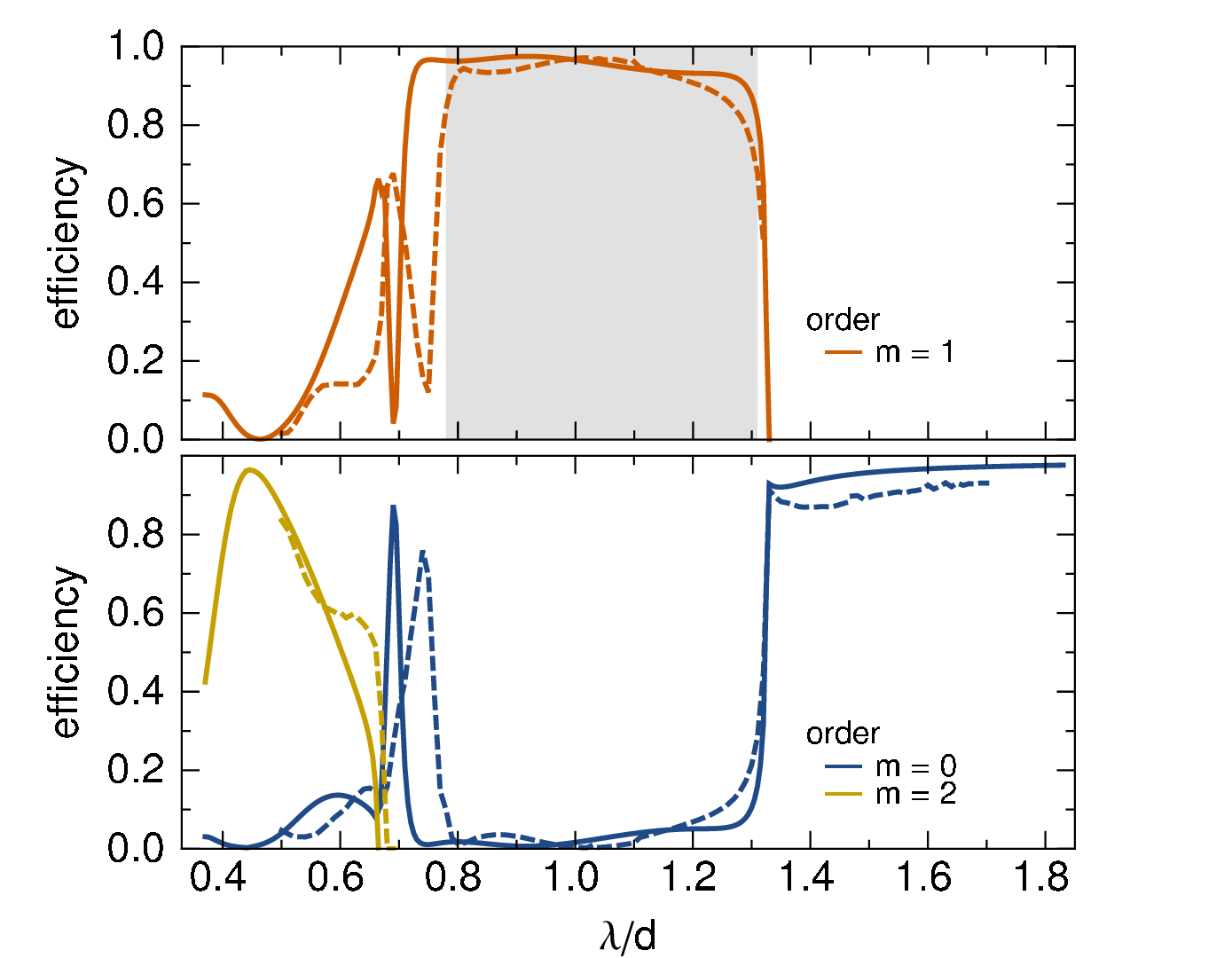}
	 \caption{Efficiency curves of a gold-plated reflection grating for radiation polarized perpendicular to the grooves, computed with the codes {\it PCGrate} (solid curves) and {\it GSolver} (dashed curves). Top graph: first-order diffraction $m=1$. The wavelength range $0.78<\lambda/d<1.31$ marked by the shaded area is used as a basis for the spectrometer layout providing an almost flat efficiency. Bottom graph: diffraction orders $m=0$ and $m=2$. Above $\lambda_0/d=1.33$ all radiation is directed into the zeroth order which simply means that the grating acts as a plane mirror. For small wavelengths ($\lambda/d<0.78$) all three orders $m=0,1,2$ contribute to the diffraction pattern. In order to avoid ambiguities this wavelength range must be removed by filtering the incident radiation.}
	 \label{grating-efficiency}
	\end{figure}

The efficiency of a grating in the diffraction order $m$ is defined as the ratio of diffracted light energy to incident energy. It was computed with two commercial codes ({\it PCGrate-S6.1} by I.I.G.\,Inc. and {\it GSolver V4.20c} by Grating Development Company) which are in reasonable agreement. In Fig.~\ref{grating-efficiency}, the efficiencies as a function of wavelength are shown for the diffraction orders $m=0,1,2$. In the range $0.78<\lambda/d<1.31$ first-order diffraction ($m=1$) dominates and has an high efficiency for linearly polarized radiation whose electric field is perpendicular to the grooves of the grating. This is essentially the useful free spectral range of the grating. All radiation with $\lambda>\lambda_0=1.33\,d$ is directed into the zeroth order, which means that the grating acts as a mirror. The short-wavelength range below $0.78\,d$ may cause problems because different diffraction orders overlap. Radiation in this range must be removed by a preceding grating stage to avoid confusion caused by overlapping orders.

\subsection{Multiple grating configuration}
\begin{figure}[tbp]
	 \centering
	 \includegraphics[width=0.85\columnwidth]{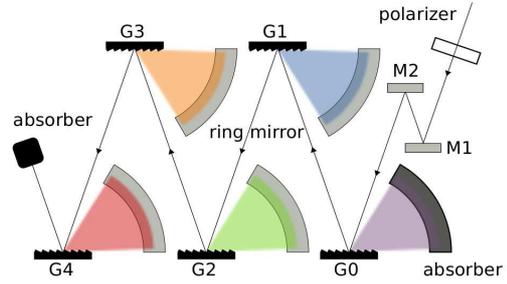}
	 \caption{Principle of the staged spectrometer equipped with five reflection gratings. Explanations are given in the  text. To avoid FIR absorption in humid air, the spectrometer is mounted in a vacuum vessel (not shown). The detector arrays mounted above the focusing mirrors are not displayed.}
	 \label{5grating-spectr}
	\end{figure}

\begin{figure}[t!]
	 \centering
	 \includegraphics[width=0.9\columnwidth]{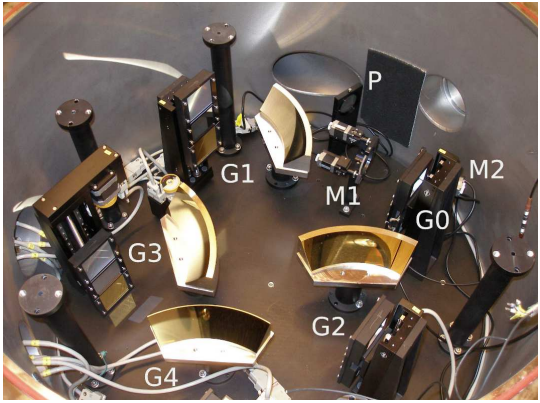}
	 \caption{Photo of the multi-grating spectrometer as mounted in a vacuum vessel attached to the CTR beam line. The detector arrays are not yet installed, hence four reflection gratings, G0 to G3, are visible. G4 is just outside the photo but its mirror can be seen. P is the polarizer, and M1, M2 are the input alignment mirrors.}
	 \label{foto-spectrometer}
	\end{figure}

\begin{figure}[h!]
	 \centering
	 \includegraphics[width=0.65\columnwidth]{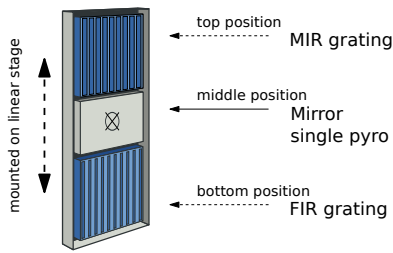}
	 \caption{The corresponding pairs of gratings are mounted on vertical translation stages with the aluminium-plated mid-infrared gratings in the upper position and the gold-plated far-infrared gratings in the lower position. Between each grating pair is either a plane mirror (for G1, G2, G3) or a pyro-electric detector (for G0 and G4) which are used for alignment purposes.}
	 \label{grating-mount}
	\end{figure}

Our spectrometer is equipped with five  consecutive  reflection gratings, G0 to G4 (see Fig.~\ref{5grating-spectr}). A photo of the spectrometer is shown in Fig.~\ref{foto-spectrometer}. Each grating exists in two variants, for the mid-infrared (MIR) and the far-infrared (FIR) regime. The parameters are summarized in Table~\ref{grating-param}. The MIR and FIR gratings are mounted on top of each other in vertical translation stages (Fig.~\ref{grating-mount}). Between each grating pair is either a mirror (for G1, G2 and G3) or a pyroelectric detector (for G0 and G4) which are needed for alignment.

                                                                                                                                                                                                                                                                                                                                                                                                                                                                                                                                                                                                           In the following we describe the far-infrared (THz) configuration, the mid-infrared configuration works correspondingly. The incident radiation is passed through a polarization filter (HDPE thin film THz polarizer by TYDEX) to select the polarization component perpendicular to the grooves of the gratings, and is then directed towards grating G0 which acts as a low-pass filter: the high-frequency part of the radiation (wavelength $\lambda<\lambda_0=44$\,\textmu m) is dispersed and guided to an absorber, while the low-frequency part ($\lambda>\lambda_0$) is specularly reflected by G0 and sent to grating G1. This is the first grating stage of the spectrometer.    Radiation in the range $\lambda_{\mathrm{min}}=45.3$\,\textmu m$\,<\lambda<\lambda_{\mathrm{max}}=77.4$\,\textmu m is dispersed in first-order and focused by the use of a ring mirror onto a multi-channel detector array, while radiation with $\lambda>\lambda_0=77.6$\,\textmu m is specularly reflected and sent to G2. The subsequent gratings work similarly and disperse the wavelength intervals $[77.0,131.5]$\,\textmu m (G2), $[140.0, 239.1]$\,\textmu m (G3), and $[256.7, 434.5]$\,\textmu m (G4).

	\begin{table}[h!] 
	\renewcommand{\arraystretch}{1.25}
	\caption[]{Parameters of the gratings. The distance between two grooves is called $d$. For light with $\lambda\ge\lambda_0$ the grating acts as a plane mirror. The mimimum and maximum wavelengths of the free spectral range for first-order diffraction are called $\lambda_{\mathrm{min}}$ and $\lambda_{\mathrm{max}}$. All dimensions are quoted in \textmu m. The coarse gratings with $d\ge\;$58.82\,\textmu m are gold-plated and were custom-made by Kugler Precision, the fine gratings with $d\le\;$33.33\,\textmu m are aluminium-plated and were purchased from Newport Corporation.}
	\label{grating-param}
	\begin{center} 
	\begin{tabular}{c}
		mid-infrared configuration $5.1-43.5$\,\textmu m
	\end{tabular}
	\begin{tabular}{ccccc}
	\hline
	grating	&	$d$	& $\lambda_{\mathrm{min}}$&	$\lambda_{\mathrm{max}}$& $\lambda_{\rm 0}$ \\
	\hline
	G0	&	4.17	& -		& -		& 5.5	\\	
	\hline 
	G1  &	6.67	& 5.13	& 8.77	& 8.8	\\
	\hline 
	G2	&	11.11	& 8.56	& 14.6	& 14.7	\\
	\hline 
	G3	&	20.0	& 15.4	& 26.3	& 26.4	\\
	\hline 
	G4	&	33.33 	& 27.5	& 43.5	& -		\\
%	\hline 
	\end{tabular}
	\vspace{3mm}

	\begin{tabular}{c}
		far-infrared configuration $45.3-434.5$\,\textmu m
	\end{tabular}
	\begin{tabular}{ccccc}
	\hline
	grating	&	$d$ &	$\lambda_{\mathrm{min}}$& $\lambda_{\mathrm{max}}$& $\lambda_{\rm 0}$ \\
	\hline
	G0	& 33.33	& -		& -		& 44	\\	
	\hline 	
	G1  & 58.82	& 45.3	& 77.4	& 77.6	\\
	\hline 
	G2	& 100.0	& 77.0 	& 131.5	& 132	\\
	\hline 
	G3	& 181.8	& 140.0	& 239.1	& 240	\\
	\hline 
	G4	& 333.3	& 256.7	& 434.5	& -		\\
%	\hline 
	\end{tabular}
	\end{center} 
	\end{table}

\subsection{Imaging system}
\begin{figure}[h!]
	 \centering
	 \includegraphics[width=0.8\columnwidth]{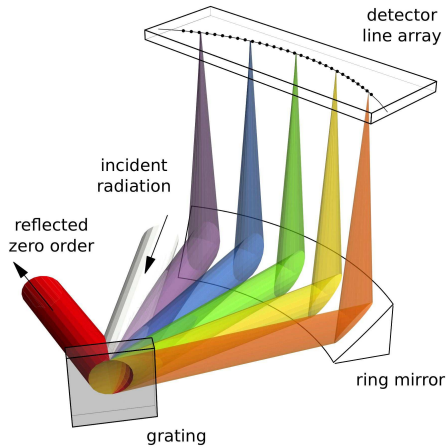}
	 \caption{Arrangement of the grating, the ring mirror and the array of 30 pyroelectric detectors. The light dispersion and focusing have been computed with a ray tracing code. For clarity only 5 of the 30 wavelength channels are shown.}
	 \label{Fig-Spektr}
	\end{figure}

For each grating, the first-order diffracted radiation is focused by a ring mirror of parabolic shape~\cite{Hossein} onto an array of 30 pyroelectric detectors which are arranged on a circular arc (Fig.~\ref{Fig-Spektr}).

\begin{figure}[h!]
	 \centering
	 \includegraphics[width=0.6\columnwidth]{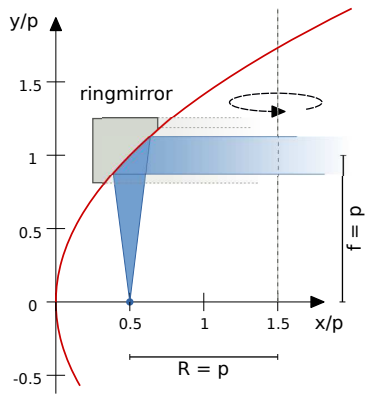}
	 \caption{The ring mirror is obtained by rotating a section of the parabola $y=\sqrt{2px}$ about a vertical axis at $x=3p/2$. The gold plated mirrors are visible in Fig.~\ref{foto-spectrometer}.}
	 \label{Ringspiegel}
	\end{figure}

The grating spreads the first-order dispersed  radiation over an  angular range from 27$^\circ$ to 80$^\circ$. To focus the light onto a cicular arc passing through the 30-channel  detector array, a ring-shaped parabolic mirror has been designed with an angular acceptance of 60$^\circ$.  The ring mirror design is shown in Fig.~\ref{Ringspiegel}. The indicated  35\,mm wide section of the parabola $y=\sqrt{2 p x}$  is rotated  about a vertical  axis at $x=3p/2$. The radius of curvature is $R=p=150$\,mm. The grating is located in the center of the ring mirror which deflects the radiation by 90$^\circ$ and has a  focal length of $f=p=150$\,mm.

\subsection{Detector array}
\begin{figure}[tbp]
	\centering
	\includegraphics[width=0.65\columnwidth]{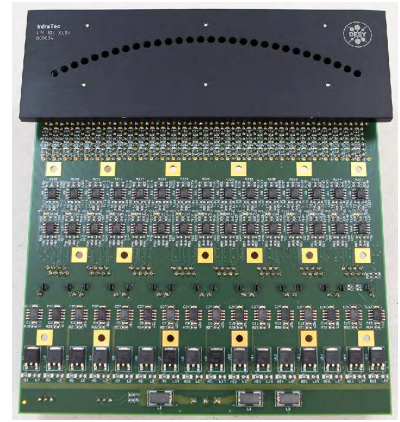}
	\caption{Photo of the 30-channel pyroelectric detector array with integrated preamplifiers and twisted-pair line drivers.}
	\label{detector-array}
	\end{figure}

The first-order diffracted radiation is recorded in a specially designed detector array equipped with 30 pyroelectric detectors (Fig.~\ref{detector-array}) which are arranged on a circular arc covering 57$^\circ$ and thus matching the acceptance of the focusing ring mirrors.

A critical component of the broadband single-shot spectrometer is a detector featuring high sensitivity over the entire THz and infrared  regime. Only bolometric devices, responding to the deposited radiation energy through a temperature rise, can cover such a wide wavelength range. The simplest and widely used bolometric detectors are pyroelectric crystals. Although inferior to cryogenic detectors in sensitivity, they have many advantages: they are very compact, comparatively inexpensive, and do not require an entrance window.
   
\begin{figure}[htbp]
	\centering
	\includegraphics[width=0.85\columnwidth]{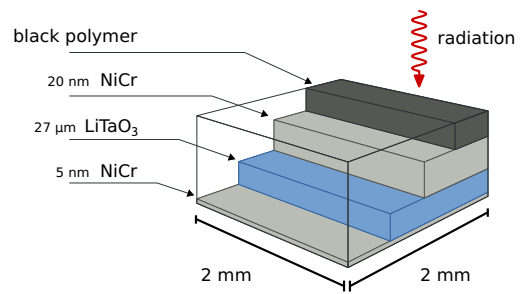}
	\caption{Layout of the pyroelectric detector LIM-107-X005.}
	\label{pyro-detector}
	\end{figure}

A special pyroelectric detector has been developed by a industrial company (InfraTec) according to our specification. This sensor possesses sufficient sensitivity for the application in a coherent radiation spectrometer and has a fast thermal and electrical response. The layout of the detector is shown in Fig.~\ref{pyro-detector}. It consists of a 27\,\textmu m thick lithium tantalate (LiTaO$_3$) crystal with an active area of 2\,$\times$\,2\,mm$^2$. The front surface is covered with a fairly thick NiCr electrode (20\,nm instead of 5\,nm). The backside electrode is a NiCr layer of only 5\,nm thickness instead of the conventional thick gold electrode. To enhance the radiation absorption below 100\,\textmu m the front electrode is optionally covered with a black polymer layer.

\begin{figure}[t!]
	\centering
	\includegraphics[width=0.9\columnwidth]{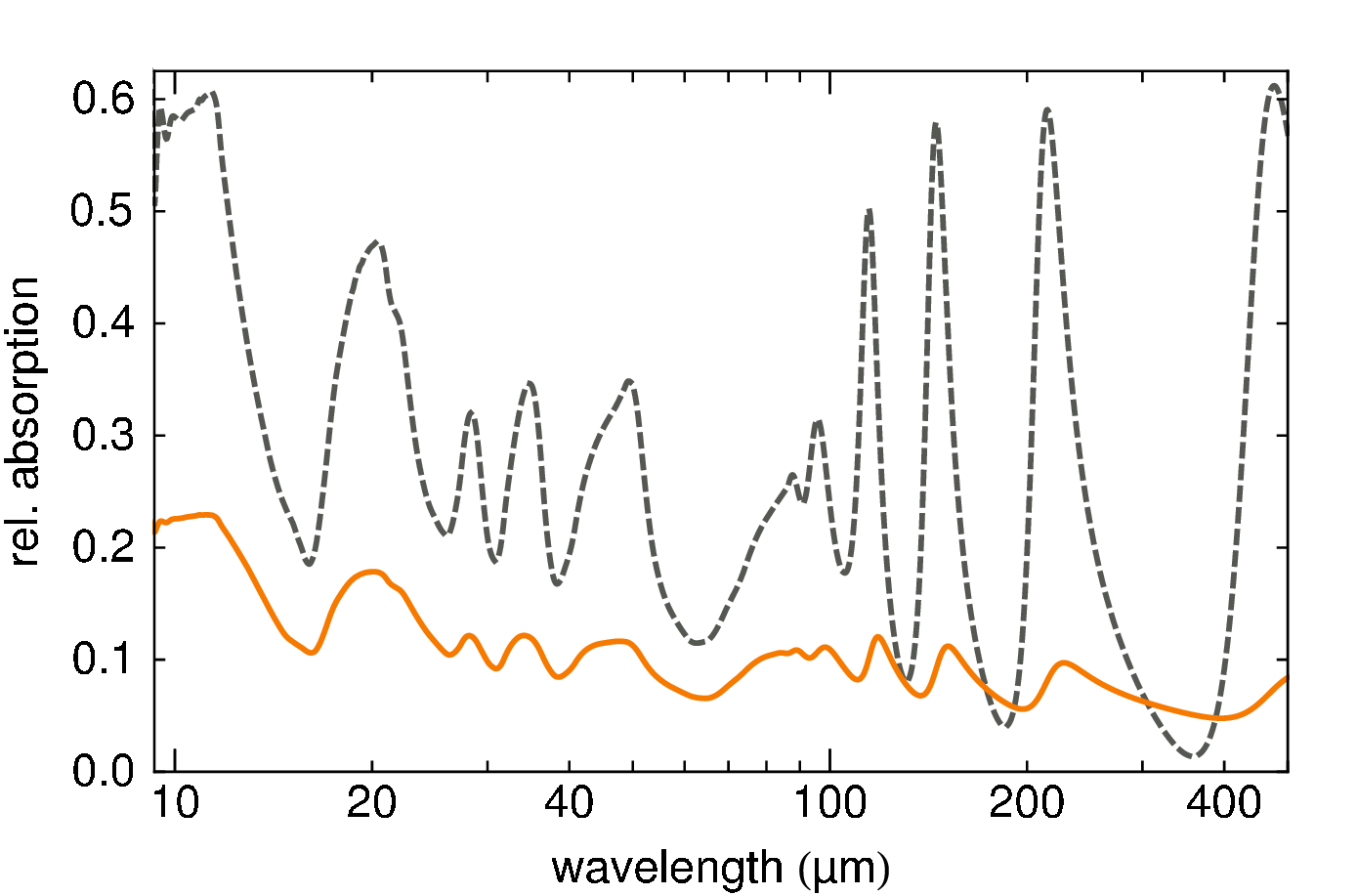}
	\caption{Computed infrared absorption as a function of wavelength. Dashed curve: 27\,\textmu m LiTaO$_3$ detector with standard coatings: a 5\,nm NiCr layer at the front surface and a thick gold layer at the back surface. Solid curve: 27\,\textmu m LiTaO$_3$ detector with optimized coatings for minimizing internal reflections (20\,nm NiCr at front surface and 5\,nm NiCr at back surface).}
	\label{LiTaO3-surface-layers}
	\end{figure}

The combination of a thick front electrode and a thin backside metallization greatly reduces internal reflections which are the origin of the strong wavelength-dependent efficieny oscillations observed in conventional pyroelectric detectors. The beneficial effect of the novel layer structure is illustrated in Fig.~\ref{LiTaO3-surface-layers}. Efficiency variations for the new pyroelectric detector (solid curve) are strongly suppressed compared to a conventional detector (dashed curve) leading to superior performance characteristics in case of spectroscopic applications. A drawback is the lower overall efficiency, but this is of minor importance in our case as there is ample coherent radiation intensity at FLASH.

	\begin{figure}[b!]
	\centering
	\includegraphics[width=0.9\columnwidth]{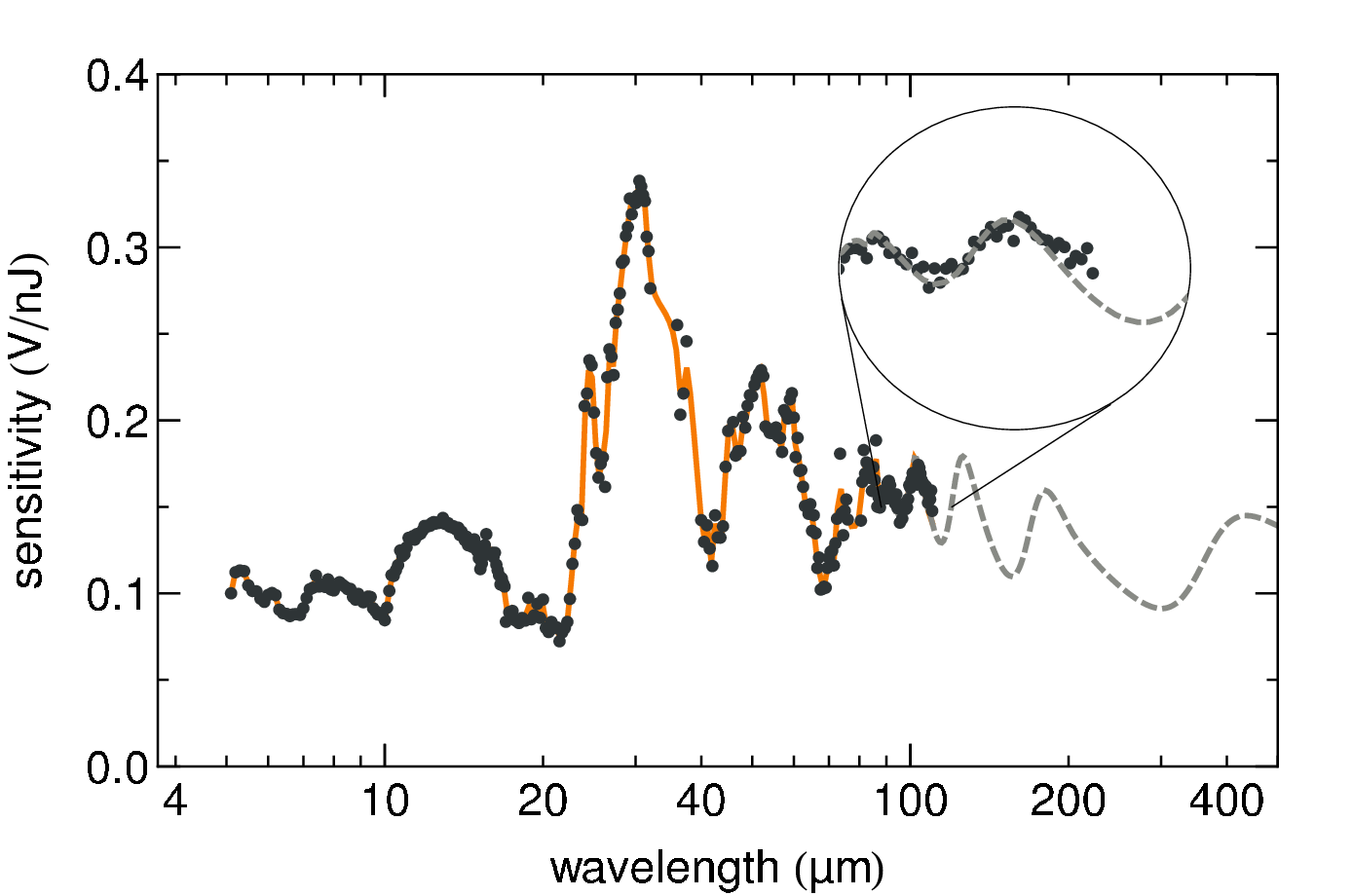}
	\caption{Calibration of the pyroelectric detector equipped with a charge-sensitive preamplifier and subsequent Gaussian filter amplifier (Fig.~\ref{amplifier-electronics}). The sensitivity is plotted as function of wavelength. Below 110\,\textmu m, the experimental data from FELIX (dots) are shown together with an interpolating curve. For longer wavelengths, a calculated curve for the novel coating arrangement (see Fig.~\ref{LiTaO3-surface-layers}) is shown as dashed line. It is normalized to the measured data in the range 90 to 110\,\textmu m. Below 90\,\textmu m the properties of the black coating are poorly known hence a meaningful comparison with the model calculation is not possible.}
	\label{pyro-sensitivity}
	\end{figure}

The pyroelectric detector is connected to a charge-sensitive preamplifier (Cremat CR110) followed by a Gaussian filter amplifier (CR220) with 4\,\textmu s time constant. This combination has been calibrated~\cite{Behrens} with infrared pulses of selectable wavelength at the infrared FEL user facility FELIX~\cite{FELIX} in the wavelength range from 5 to 110\,\textmu m. The measured sensitivity as a function of wavelength is shown in Fig.~\ref{pyro-sensitivity}. While the wavelength dependence of the sensitivity has been determined rather accurately, the absolute calibration is uncertain by at least 50\%. Two different power meters were available to measure the total power in the FELIX test beam. They agreed in the wavelength dependence but the absolute power readings differed by a factor of two.

The black polymer coating introduces a thermal time constant of the detector response of typically 30\,\textmu s. Its contribution to the overall sensitivity can be studied by recording the electric pulse shape directly after the preamplifier. It turned out that the coating is basically transparent for wavelengths above 90\,\textmu m. This allows to extrapolate the sensitivity curve to longer wavelengths where no data exists by using the computed absorption for the bare sensor shown in Fig.~\ref{LiTaO3-surface-layers}.

	\begin{figure}[htbp]
	\centering
	\includegraphics[width=0.9\columnwidth]{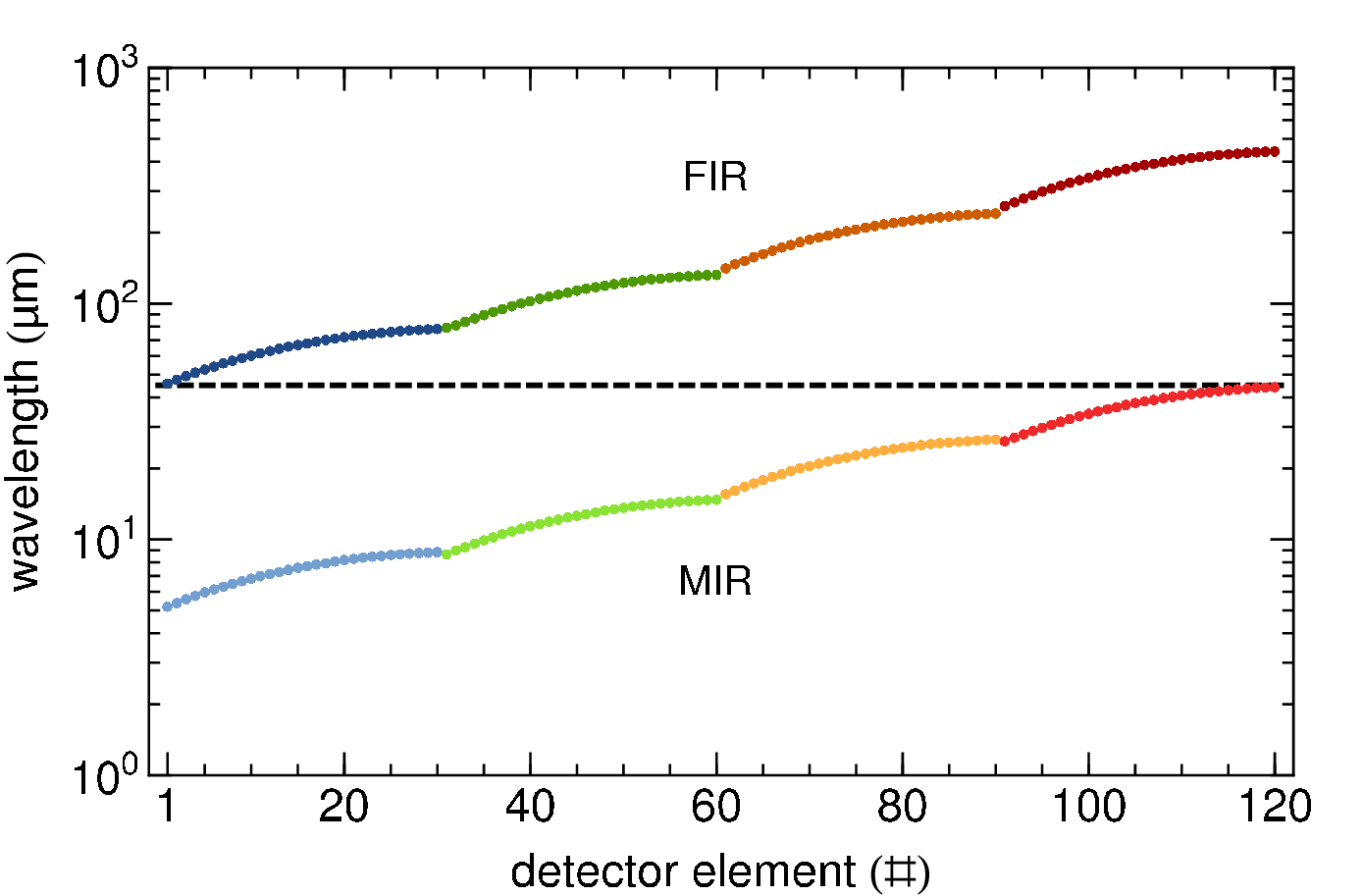}
	\caption{Wavelength vs. detector number for the mid-infrared (MIR) and the far-infrared (FIR) configuration of the multichannel spectrometer.}
	\label{wavelength-vs-detector}
	\end{figure}

For a given spectrometer configuration, mid-infrared (MIR) or far-infrared (FIR), the dispersed radiation is recorded in 120 parallel detectors. The wavelength as a function of detector number is plotted in Fig.~\ref{wavelength-vs-detector}. The MIR spectrometer covers the wavelength range from 5.1\,\textmu m to 43.5\,\textmu m, the FIR spectrometer covers the range from 45.5\,\textmu m to 434.5\,\textmu m.

	\begin{figure*}[t]
	\centering
	\includegraphics*[width=0.95\textwidth]{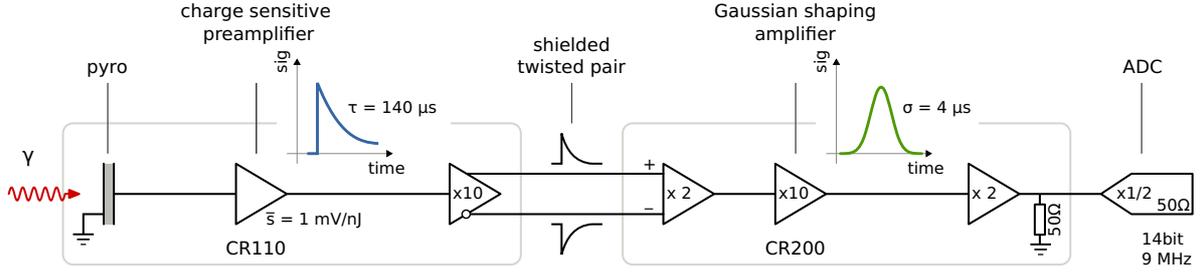}
	\caption{Electronics diagram. The preamplifier and the twisted-pair line driver amplifier are mounted on the electronics board inside the vacuum vessel (see Fig.~\ref{detector-array}). Line receiver, Gaussian shaping amplifier and ADC are located outside the vacuum vessel.}
	\label{amplifier-electronics}
	\end{figure*}

All 30 pyroelectric sensors have the same angular acceptance $\Delta \beta_1= w/f$ where $w=2\,$mm is the width of the sensor and $f=150\,$mm the focal length of the ring mirror. However, the wavelength intervall $\Delta \lambda$ subtended by a sensor depends strongly on its angular position. From Eq.~(\ref{grating-equation}) it follows with $m=1$
	\begin{equation}
	\Delta \lambda=d\, \cos{\beta_1} \cdot \Delta \beta_1
	\label{Delta-lambda(beta)}
	\end{equation}
hence $\Delta \lambda$ drops rapidly with increasing detector number since the angle $\beta_1$ grows from $\beta_1=27^\circ$ at detector 1 to $\beta_1=80^\circ$ at detector 30.

The complete electronics diagram is depicted in Fig.~\ref{amplifier-electronics} where also the pulse shape at various stages can be seen. The commercial charge-sensitive preamplifier CR110 generates pulses with a rise time of 35\,ns and a decay time of 140\,\textmu s. A Gaussian shaping amplifier (Cremat CR200 with 4\,\textmu s shaping time) is used to optimize the sigal-to-noise ratio. This is adequate for the application at the CTR beamline~\cite{Casalbuoni} where the repetition rate is only 10\,Hz. The shaped signals are digitized with 120 parallel ADCs with 9\,MHz clock rate, 14-bit resolution and 50\,MHz analog bandwidth. A fast version of the electronics with 250\,ns shaping time, allowing 1\,MHz repetition rate, is used for an identical spectrometer inside the FLASH tunnel. This system is presently being commissioned.  
	
\subsection{Spectrometer alignment and computed response function}
Within each spectrometer stage the  alignment of the grating, the ring mirror and the detector line array was done with a laser beam. The overall alignment of the spectrometer was carried out in situ with coherent transition radiation emerging from the CTR beamline~\cite{Casalbuoni} which will be described in the next section. For this purpose the vertical translation stages were moved to their mid-position (see Fig.~\ref{grating-mount}). The two input alignment mirrors M1 and M2 (see Fig.~\ref{5grating-spectr}) can be remotely adjusted in two orthogonal angles. Varying these four independent angles the radiation was centered on the pyroelectric sensors mounted on the G0-stage and the G4-stage.

	\begin{figure}[b!]
	\centering
	\includegraphics[width=0.9\columnwidth]{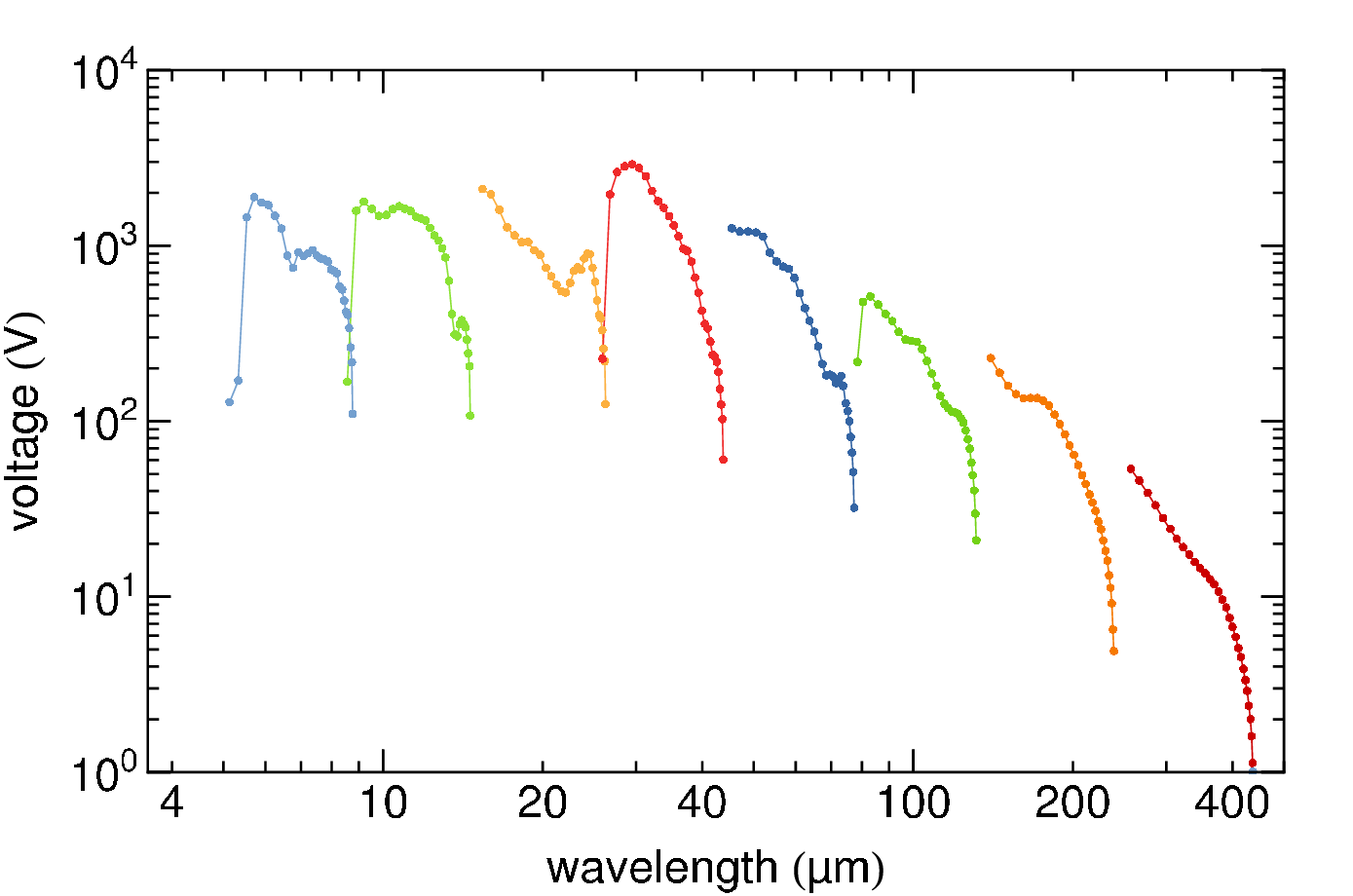}
	\caption{Computed overall response function of the multichannel spectrometer as installed at the CTR beamline at FLASH and the beam parameters given in the text. The expected voltage at the 14-bit ADCs is plotted as a function of wavelength. The sawtooth-like structure is caused by the variation of wavelength acceptance within each detector array, described by Eq.~(\ref{Delta-lambda(beta)}).}
	\label{spectrometer-response}
	\end{figure}

	\begin{figure*}[htbp]
	\centering
	\includegraphics*[width=0.95\textwidth]{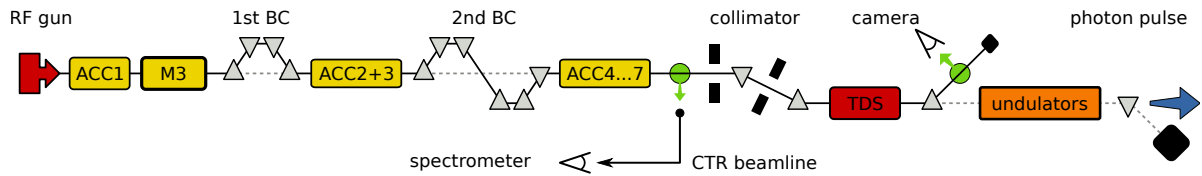}
	\caption{Schematic layout of the free-electron laser FLASH. Shown are the components which are important for the present experiment: radio-frequency (RF) gun, the seven acceleration modules ACC1 to ACC7, the third-harmonic cavity module M3, and the two magnetic bunch compressors BC. The transition radiation screen and the diamond window of the coherent transition radiation (CTR) beamline are installed behind ACC7. The radiation is guided to a spectrometer outside the accelerator tunnel. The transverse deflecting microwave structure TDS is mounted in front of the undulator magnets. }
	\label{FLASH_overview}
	\end{figure*}

The computed overall response function of the multichannel spectrometer as installed at the CTR beamline is shown in Fig.~\ref{spectrometer-response}. The power transmission of the beamline depends on the wavelength but in addition on the emission characteristics of the radiation source. A dedicated {\small Mathematica} code, {\it THzTransport}, was developed to model the generation of coherent transition radiation by an infinitely short electron bunch (longitudinal form factor $F_\ell\equiv1$) with an rms radius of $\sigma_r=200$\,\textmu m and a charge of 1\,nC. The same code was applied to simulated the transport of the radiation through the CTR beamline, taking near-field diffraction effects into account. A detailed description of the numerical procedures can be found in~\cite{Casalbuoni}. Additionally, the following wavelength-dependent effects were taken into consideration: transmission of the polarization filter, grating efficiencies (Fig.~\ref{grating-efficiency}), pyroelectric detector sensitivity (Fig.~\ref{pyro-sensitivity}), and wavelength acceptance $\Delta\lambda$ of the sensors as a function of grating constant $d$ and sensor number according to Eq.~(\ref{Delta-lambda(beta)}). 

The Fourier transformation of an infinitely short electron bunch yields a flat frequency spectrum, and because of $\omega=2\pi c /\lambda, d\omega=-2\pi c /\lambda^{2}~d\lambda$, a wavelength spectrum proportional to $1/\lambda^2$. The angular distribution of transition radiation from a screen of finite size together with diffraction losses in the  beam line and in the focusing of the radiation onto the small pyroelectric detectors contribute also to the drop of the spectral intensity towards large wavelengths. In the small-wavelength regime the response function levels off. This is due to the finite transverse electron beam size.

\section{Commissioning of the spectrometer and benchmarking results}
\label{results}	

\subsection{The FLASH facility}
The soft X-ray FEL facility FLASH is schematically shown in Fig.~\ref{FLASH_overview}. A detailed description of the design criteria and the layout can be found in~\cite{FEL-book,Schreiber}, here we give only a short overview. The electron source is a laser-driven photocathode mounted in a 1.3\,GHz copper cavity. The main components of the linear accelerator are seven 1.3\,GHz superconducting acceleration modules (ACC1 to ACC7) and two magnetic bunch compressors BC. The electron gun generates trains of electron bunches at a repetition rate of 10\,Hz, each train comprising up to 800 bunches with a time spacing of 1\,\textmu s. The electron bunches leaving the gun have an energy of 4.5\,MeV and an adjustable charge of 0.02 to 1.5\,nC. The energy is boosted to about 150\,MeV in module ACC1. The module is operated at an off-crest phase in order to induce an energy gradient along the bunch axis that is needed for longitudinal bunch compression in the magnetic chicanes. The recently installed third-harmonic cavity module M3~\cite{ACC39} linearizes the energy chirp and is essential for producing bunches of variable length. Modules ACC2 and ACC3 accelerate the electron bunches to about 450\,MeV, and the final compression takes place in the second BC. The acceleration modules ACC4 to ACC7 boost the energy up to a maximum of 1.2\,GeV. The undulator magnet consists of six 4.5\,m long segments. Two collimators in combination with a magnetic deflection protect the permanent magnets of the undulator against radiation damage by beam halo.

\subsection{CTR beam line}
To facilitate frequency domain experiments at FLASH, an ultra-broadband optical beamline for transition radiation (TR) was built which transports electromagnetic radiation ranging from 0.2\,THz up to optical frequencies and permits spectroscopic measurements in a laboratory outside the accelerator tunnel~\cite{Casalbuoni}. The radiation is produced on an off-axis screen by single electron bunches that are picked out of the bunch train by a fast kicker magnet. The TR screen is a 350\,\textmu m thick polished silicon wafer with a 150\,nm thick aluminium coating on the front surface. The screen is tilted by 45$^\circ$ with respect to the electron beam direction, hence the radiation is emitted perpendicular to the beam. It leaves the electron beam vacuum chamber through a 0.5\,mm thick diamond window with diameter of 20\,mm. The transmission is 71\% (caused by reflection losses at the two surfaces) for wavelength between 400\,nm and 3\,mm, except for a narrow absorption band of diamond between 4 and 7\,\textmu m where the transmission drops down to 40\%. The radiation is guided through a 20\,m long evacuated beam line equipped with 6 focusing mirrors to a vacuum vessel outside the accelerator tunnel which houses the spectrometer or other diagnostic devices.

	\begin{figure}[b!]
	\centering
	\includegraphics[width=0.9\columnwidth]{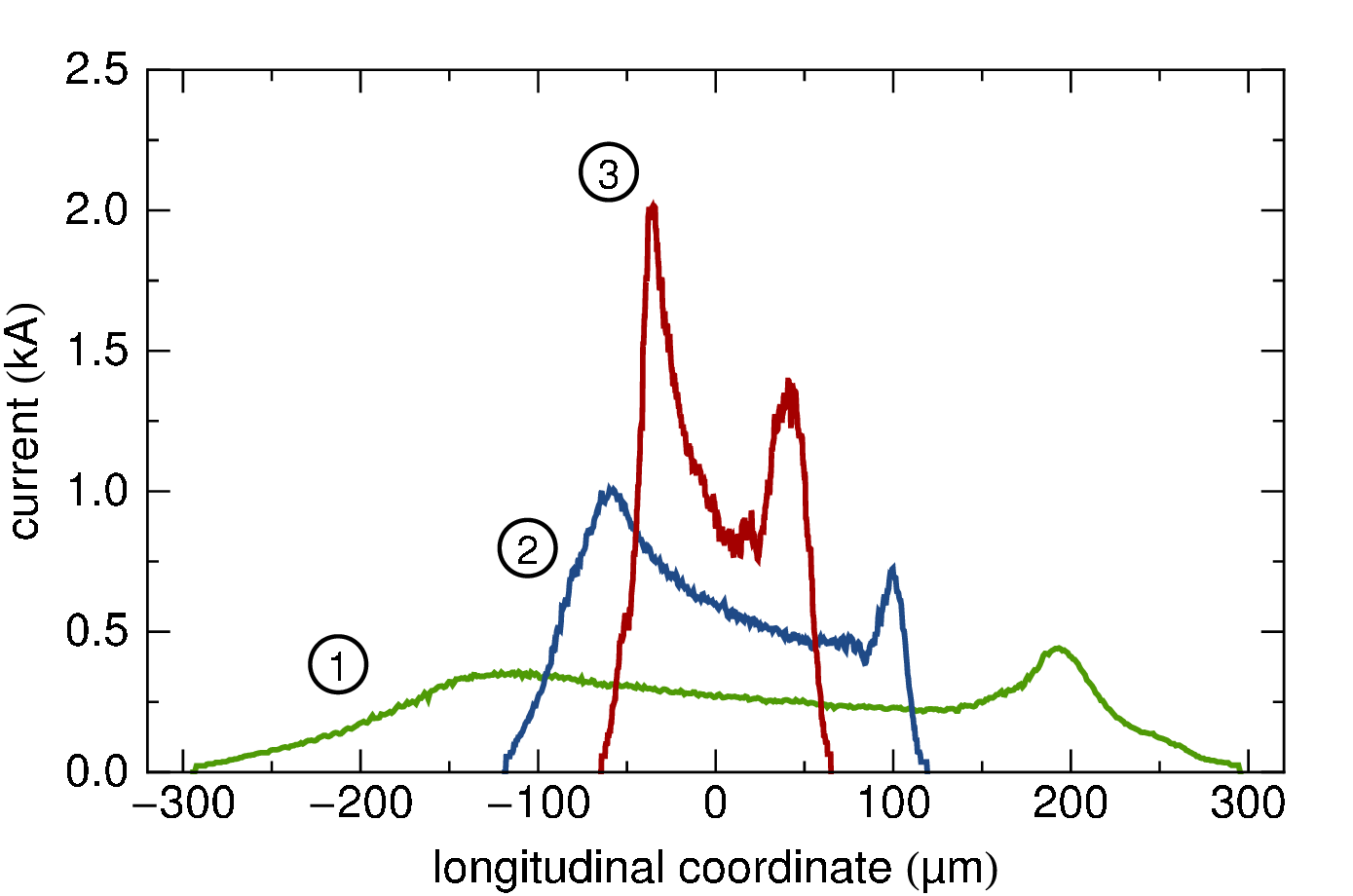}
	\caption{Longitudinal profiles of three different electron bunch shapes in the FLASH linac as measured with the transverse deflecting microwave structure.}
	\label{bunchprofile}
	\end{figure}

\subsection{Test measurements with bunches of known shape}
To verify the performance of the spectrometer, test measurements were carried out with bunches of different length and structure. The longitudinal profiles of the electron bunches were determined with high accuracy using the transverse deflecting microwave structure TDS mounted in front of the undulator (see Fig.~\ref{FLASH_overview}). Three different bunch shapes with total lengths between 100 and 400\,\textmu m were realized by a proper choice of the RF phases in the acceleration modules ACC2 and ACC3. To reduce statistical fluctuations, 40 bunches were recorded with the TDS setup for each shape. Two different sweep directions of the TDS were applied to minimize systematic errors in the determination of the longitudinal charge distribution.
 
	\begin{figure}[t!]
 	 \centering
	 \includegraphics[width=0.9\columnwidth]{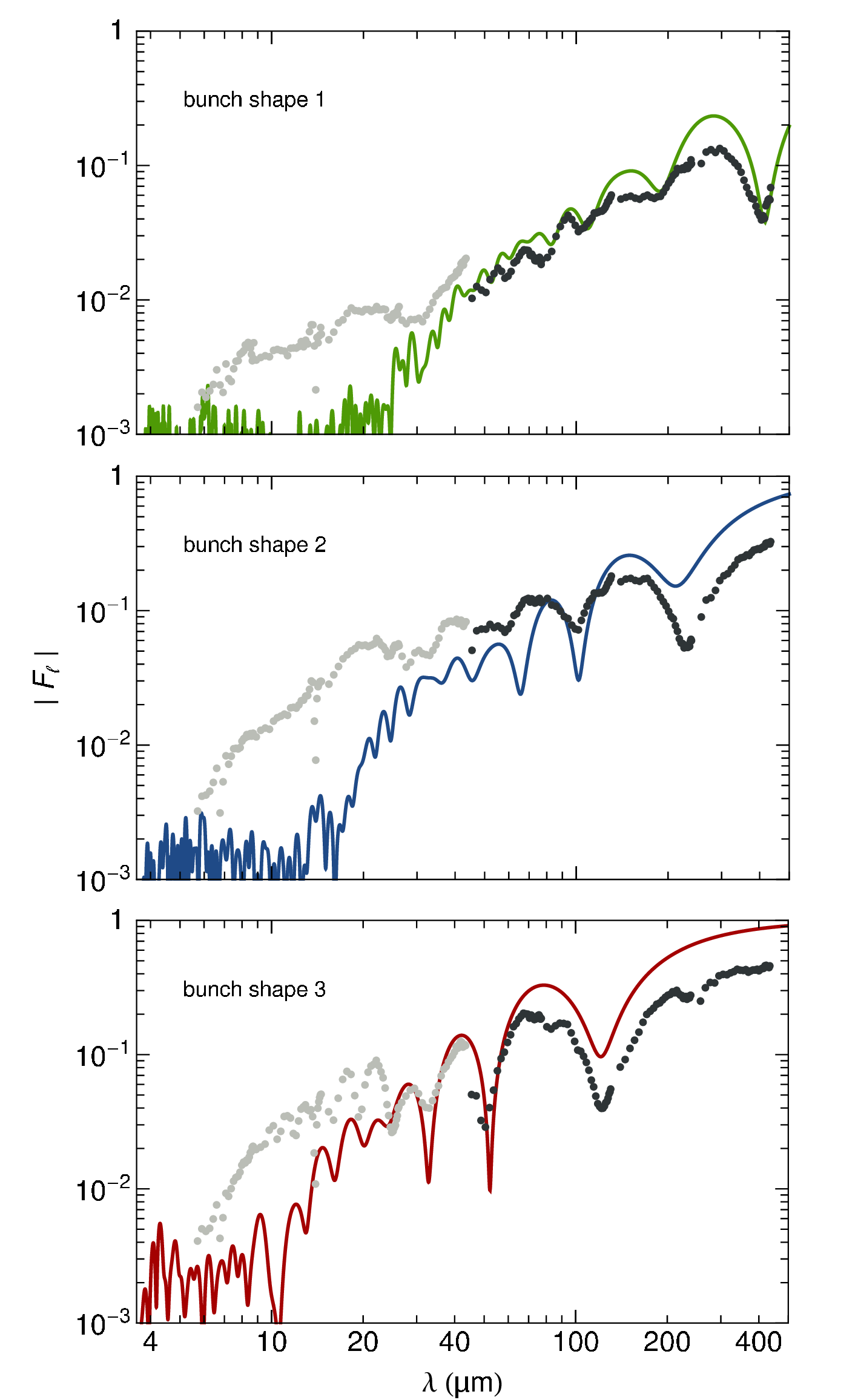}
 	 \caption{The magnitude $|F_{\ell}(\lambda)|$ of the longitudinal form factors of the bunch shapes shown in Fig.~\ref{bunchprofile}.  Solid curves: form factors computed by Fourier transformation from the TDS longitudinal bunch profiles. Dotted curves: form factors derived from the spectroscopic measurements (black dots: FIR configuration, gray dots: MIR configuration).}
	\label{bunchspectrum}
	\end{figure}

The averaged longitudinal profiles are shown in Fig.~\ref{bunchprofile}. The form factors were computed by Fourier transformation of the averaged longitudinal profiles, Eq.~(\ref{eq2}). They are shown as solid curves in Fig.~\ref{bunchspectrum}.

The rms time resolution of the TDS system depended on the total bunch length to be covered and varied between ${\sigma_t=27}$\,fs for the shortest bunches and ${\sigma_t=43}$\,fs for the longest ones. This finite time resolution of the TDS system leads to a suppression of the form factor towards small wavelengths. The suppression factor is $\exp(-\lambda_{\rm cut}^2/\lambda^2)$ with the ``cutoff wavelength'' ${\lambda_{\rm cut}=\sqrt{2}\,\pi\,c\,\sigma_t\,}$; here $c$ is the speed of light. For the three bunch shapes shown in Fig.~\ref{bunchprofile}, the cutoff wavelengths are 58, 40 and 36\,\textmu m, respectively.

For all three bunch shapes spectroscopic measurements were carried out on 200 bunches each. The form factors derived from the measured spectra are depicted as dotted curves in Fig.~\ref{bunchspectrum}. For all three bunch shapes, there is an impressive overall agreement between the form factors derived by the two complementary methods. In the short-wavelength region below 20\,\textmu m the spectroscopically determined form factors are generally higher than the formfactors derived from the TDS measurements. This is evidence for the presence of a fine structure which cannot be resolved by the TDS system in its present configuration. This is a convincing demonstration of the capabilities of the multi-channel spectrometer. 

	\begin{figure}[b!]
	 \centering
	 \includegraphics[width=0.95\columnwidth]{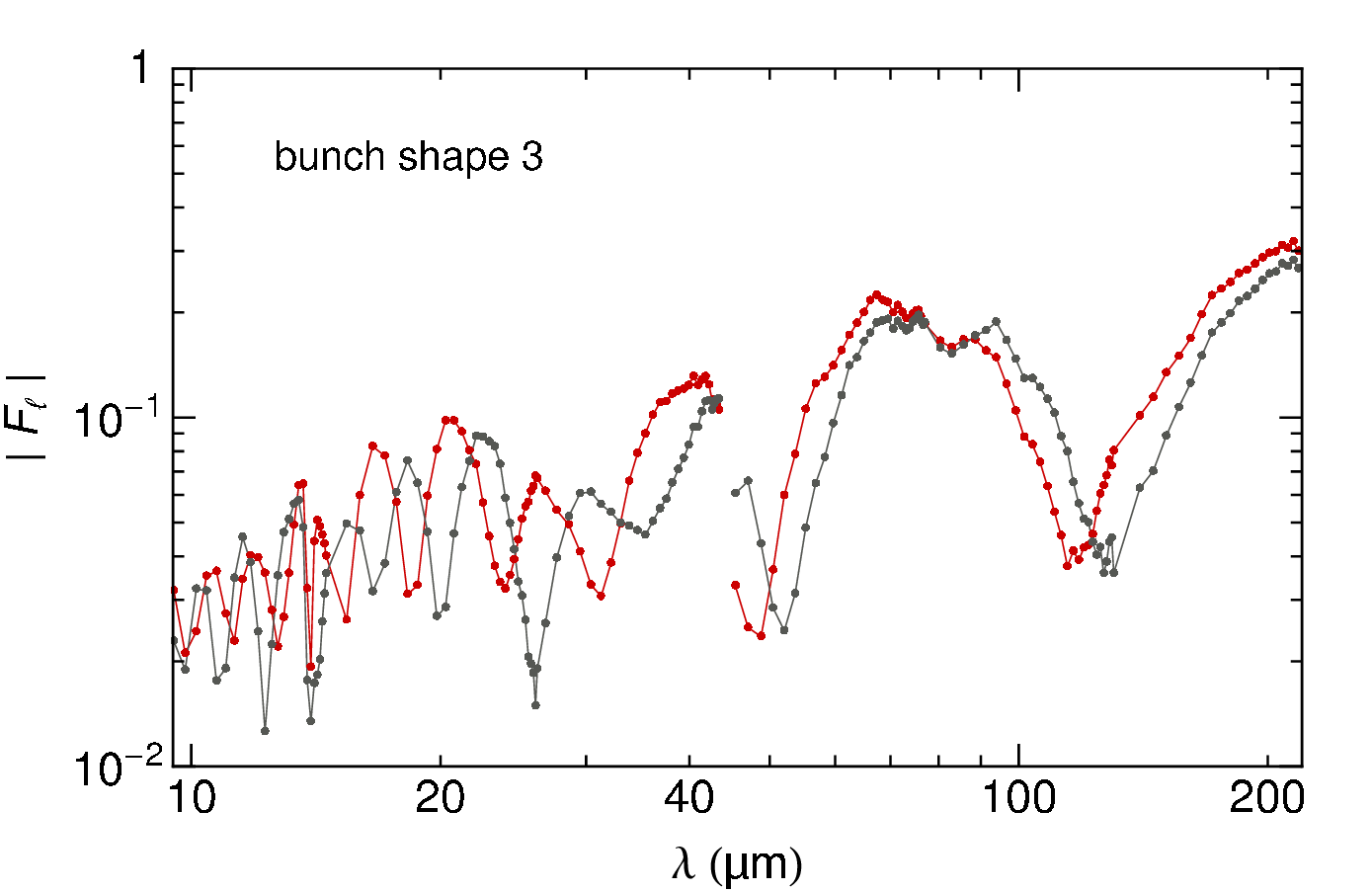}
 	 \caption{The form factor $|F_{\ell}(\lambda)|$ of two successive bunches, derived from two single-shot spectroscopic measurements.}
	 \label{two-shots}
	\end{figure}
	
The importance of the single-shot capability of the spectrometer is demonstrated in Fig.~\ref{two-shots}. The spectroscopically determined form factors of two successive bunches, having nominally the same shape, exhibit both significant structures which however are shifted against each other. This shows that there exist substantial shot-to-shot fluctuations even if the accelerator parameters are kept as constant as technically feasible. Averaging over several bunches in either system will wash out many of the structural details. Hence a perfect agreement between TDS and spectrometer data cannot be expected. With our present setup at the FLASH linac it is not possible to measure exactly the same bunch with both instruments.

\section{Summary}
A multi-channel single-shot spectrometer for the THz and infrared range has been developed and successfully commissioned with coherent transition radiation from electron bunches of known shape. The spectrometer covers a factor of 10 in wavelength with parallel readout. With two sets of remotely interchangeable gratings the wavelength ranges from 5.1\,-\,43.5\,\textmu m and 45.3\,-\,434.5\,\textmu m can be analyzed. Very good agreement is found between the spectroscopic longitudinal bunch form factors and the form factors derived from time-domain measurements using a transverse deflecting microwave structure. This is a proof of the high performance of the broadband spectrometer.

The unique single-shot capability of the multi-channel spectrometer will allow to study and monitor electron bunch profiles in great detail and will help to control the bunch compression process in view of an optimized FEL performance of the facility.

\vspace{3mm}

We thank our engineers Kai Ludwig (mechanical design) and Petr Smirnov (readout electronics) for their important contributions of the spectrometer. An essential accelerator component for the benchmarking to the spectrometer was the third-harmonic cavity module M3 which was constructed at Fermilab and commissioned at DESY under the leadership of Helen Edwards. We want to thank Helen Edwards for this important contribution to the present experiment and moreover for her careful reading of the manuscript and  valuable comments.



%% References


%\bibliographystyle{model1a-num-names}



\end{document}